\documentstyle[epsfig,prd,aps]{revtex}
\def\tt{\tau}
\begin{document}

\thispagestyle{plain}
\setcounter{page}{1}

\title{Does parton saturation at high density explain hadron
multiplicities at RHIC ?}

\author{R.~Baier$^a$, A.H.~Mueller$^b$, D.~Schiff$^c$, and D.T.~Son$^d$}

\address{
$^a$ Fakult\"at f\"ur Physik, Universit\"at Bielefeld, 
D-33501 Bielefeld, Germany\\
$^b$ Department of Physics, Columbia University, New York, NY 10027, USA\\
$^c$ LPT, Universit\'e Paris-Sud, B\^atiment 210, F-91405 Orsay, France\\
$^d$ Institute for Nuclear Theory, University of Washington,
Seattle, WA 98195-1550,  USA}

\maketitle

\begin{abstract}
We discuss the recent claim  that hadron multiplicities
measured at RHIC energies are {\sl directly} described in terms of
gluon degrees of freedom  fixed from the initial conditions
of central heavy ion collisions. The argument is  based on the parton
saturation scenario expected to be valid at high parton densities
and on the assumption of  conserved gluon number.
Alternatively we conjecture that "bottom-up" equilibration before
hadronization modifies this picture, due to nonconservation of the number
of  gluons.
\end{abstract}


\vspace{1.3cm}


At RHIC and LHC energies in the central rapidity region
of heavy ion collisions a high density of energy is deposited
mainly in the form of gluons.
In recent  papers by Kharzeev et al. \cite{Nardi,Levin,Kharzeev,Kreview} 
(hereafter referred to as KLN) it is assumed 
 that the initial gluon density 
determines the number of produced (charged) particles, i.e. 
that there is a \underbar{direct}
correspondence between the number of partons in the initial state and the 
number of particles in the final state.
The argument is formulated in the framework of the saturation scenario
\cite{Gribov,Qiu,Blaizot,MV,Krasnitz},
which determines the initial gluon distribution
inside the colliding nuclei, and
where the characteristic momentum scale is the hard scale
 $Q_s >> \Lambda_{QCD}$.

Based on this scenario the initial gluon multiplicity
immediately after the high energy nuclear collisions can be calculated
in the McLerran-Venugopalan model \cite{MV,Krasnitz,McLerran,Mueller}.
At $\tau_0 \simeq 1/Q_s $ 
this  initial gluon  density is 
\cite{Mueller},
\begin{equation}
n_{hard}(\tau_0) = c~  \frac{(N_c^2-1) Q_s^3}{4 \pi^2 N_c \alpha_s~ (Q_s \tau_0)} \,  .
\label{dens}
\end{equation}

In \cite{Nardi,Levin,Kharzeev} this relation is used to obtain
the hadron multiplicities in A-A collisions, which are then  compared
with the corresponding RHIC data \cite{Nagle,Back,Adcox,Bearden,Adler}.

In this note we propose a possibly more realistic description, which 
departs from taking the initial condition as the only one determining
 ingredient for the final hadronic state. This description is based on the 
``bottom-up'' scenario \cite{Baier1,Son,Mueller3},
which leads to thermalization of the gluons 
produced after the collision.
In contrast to KLN, this  scenario  stresses the importance of branching
processes of gluons which may  allow for a short enough equilibration time,
at least at high energies. As a consequence the number of gluons is
increasing between initial and equilibration times.
In this context let us  note that already in \cite{KrasnitzV} a 
phenomenological factor $\kappa_{inel} > 1$ is introduced
accounting for gluon number changing processes which may occur at late times
beyond when the classical approach is applicable.

In the following we investigate in the ``bottom-up'' picture the resulting
hadron multiplicities and compare with the predictions 
of the KLN  model. We discuss in some detail the 
conceptual differences between these two interesting possibilities.

The arguments of KLN \cite{Nardi,Levin,Kharzeev} go through a number
of steps, which we first critically review:

i)  The number of gluons
in the initial state, and at the time
when gluons transform into hadrons,  is  assumed to be equal, which
can be true when only $2 \leftrightarrow 2$ processes are taken into account.
Only in this circumstance the measured hadron multiplicities are reflecting
{\sl directly} the initial conditions.

ii) In (\ref{dens}) the parameter  $c$ is a constant 
linking the number of gluons
in the nucleus wave function to the number of gluons 
which are freed during the collision.
It is expected to be of $O(1)$ \cite{Mueller}. An (approximate)
analytical calculation gives $c = 2~\ln{2} \simeq 1.39$ \cite{Kovchegov}.
However,  extracting  $c$ from the numerical simulation 
in \cite{Krasnitz,KrasnitzV} leads to the estimate {\footnote{This 
estimate is based on the relation
 $c= 4 \pi^2 f_N / [(N_c^2 - 1)~\ln{Q_s^2/\Lambda^2_{QCD}} ]$,
 where $f_N = 0.3$ \cite{Raju}. }}
$c \simeq 0.5$ for $Q_s = 1~GeV$.

iii) The hard saturation scale $Q_s^2$ in the case of one nucleus
has been determined to be \cite{Mueller},
\begin{equation}
Q_s^2 (\vec{s}) =
 \frac{4 \pi^2 N_c}{N_c^2 -1}~ \alpha_s (Q_s^2)~ x G(x, Q_s^2)~
 \rho_{nucleon}(\vec{s})  \, ,  
\label{nscale}
\end{equation}
where $\vec{s}$ is the impact parameter, $x G(x, Q^2)$ is the gluon
structure function in the nucleon, and $\rho_{nucleon}$ is the transverse
density of nucleons in the nucleus.
Notice that in the above equation an additional
 numerical multiplicative factor
may come as a consequence of the inherent uncertainty 
in the precise determination of the saturation momentum
outside of the McLerran-Venugopalan model \cite{MV}.
 We shall call this factor  $K$ and introduce it a little later on.
KLN generalize (\ref{nscale}) to the case of two colliding nuclei.
Let us first write
{\footnote{Notice in (\ref{nuscale}), consistently with (\ref{nscale}),
 the factor $\frac{1}{2}$ compared to 
eq.~(14) in \cite{Nardi}.
This has the consequence of lower values of $Q_s^2$
appearing in our discussion of the KLN approach.}}

\begin{equation}
Q_s^2 (\vec{s},\vec{b}) =
 \frac{4 \pi^2 N_c}{N_c^2 -1}~ \alpha_s (Q_s^2)~ x G(x, Q_s^2)
~\frac{ \rho_{part}(\vec{s},\vec{b})}{2}  \, .  
\label{nuscale}
\end{equation}
where $\rho_{part}(\vec{s},\vec{b})$ is the density of
 participating nucleons in the transverse
plane as a function of the impact parameter 
$\vec{b}$ of the A-A collision, and of $\vec{s}$ the transverse coordinate of
the produced gluon.
The factor $\frac{1}{2}$ is required by the proper definition of $Q_s^2$,
relative to one nucleus \cite{Mueller}.
Integrating the density with respect to $\vec{s}$ leads to 
\begin{equation}
\int~ d^2s ~\rho_{part}(\vec{s},\vec{b}) = N_{part} (\vec{b}) ~,
\label{npart}
\end{equation}
with $N_{part} (\vec{b}) $ the number of participants in the A-A collision
at fixed $\vec{b}$.

iv) Focusing on the distribution of freed gluons 
 $dN/{d \eta}$
for ${\eta=0}$  at given $\vec{b}$, we can write
\begin{eqnarray}\label{spect}
 \frac{d N}{d \eta} (\vec{b})  
&& = c \frac{N_c^2 -1 }{4 \pi^2 N_c} \, 
   \int~ d^2s ~  \frac{1}{\alpha_s} \,
Q_s^2(\vec{s},\vec{b})  \nonumber  \\ 
&& \simeq c  ~ xG(x, {\bar Q_s^2}) ~\frac{ N_{part} (\vec{b})}{2}
\, .
\end{eqnarray}
The corresponding charged hadron multiplicity 
for the most central collisions is then obtained as
\begin{equation}\label{Nard0}
\langle \frac{2}{N_{part}} \frac{d N_{ch}}{d\eta} \rangle 
\simeq \frac{1}{3} ~c~\left[ \ln \frac{\bar{Q^2_s} }{\Lambda^2_{QCD}} \right] \, , 
\end{equation}
using the gluon structure function (\ref{glu}) given below.
As reference energy we take $\sqrt{s} = 130~GeV$.

One should remark that in (\ref{spect})
and (\ref{Nard0}), ${\bar Q_s^2}$ shows up as an effective
average over the variable $\vec{s}$ in (\ref{nuscale}).
As a first approximation the following is used:
\begin{equation}\label{aver}
{\bar{Q_s^2}} (\vec{b}) =
 K~\frac{4 \pi^2 N_c}{N_c^2 -1}~ \alpha_s ({\bar{Q_s^2}})~ x G(x,{\bar{ Q_s^2}})
~\frac{ \rho_{part}(\vec{b})}{2}  \, .  
\end{equation}

v) Information on the gluon structure function is necessary. In the small
$x$ regime, the main feature is that it increases with $Q^2$ at fixed $x$.
Following KLN \cite{Nardi,Levin,Kharzeev}, it is reasonable to take
\begin{equation}
 x G(x, Q^2) = 0.5 ~\ln (\frac{Q^2}{\Lambda_{QCD}^2}) \, ,
\label{glu}
\end{equation}
with $\Lambda_{QCD} = 200~ MeV$, such that 
$ x G(x, Q^2) \simeq 2 $ at $Q^2 = 2 ~GeV^2$ (at $x=0.02$).

vi) The strong coupling constant is 
\begin{equation}
 \alpha_s(Q^2) \simeq  \frac{1}{\beta_0~\ln (\frac{Q^2}{\Lambda_{QCD}^2})} \, ,
\label{coupl}
\end{equation}
with $\beta_0 = (11 - 2 n_f/3)/4 \pi$
for $N_c =3$; we  take $n_f = 3$. 

vii) Using (\ref{aver}) together with the above
choices for $\alpha_s$ and $xG(x,Q^2)$, one finds 
for central $Au-Au$ collisions at RHIC (at $\sqrt{s} = 130~ GeV$), taking for the moment
$K = 1$:

 ${\bar{Q_s^2}} (\vec{b} =0) \simeq 0.63 ~GeV^2$,
using $\rho_{part} (\vec{b} =0) = 3.06 ~fm^{-2}$,

\noindent
as quoted in Table 2 by \cite{Nardi}.
For larger values of $\vec{b}$, $\bar{Q_s^2}$
becomes even smaller, e.g. for ${b} = 10~ fm$, $\bar{Q_s^2} 
\simeq 0.32~ GeV^2$ !
This casts some doubts on the applicability of this model,
even for central collisions at RHIC energies.

As already noticed, the missing factor $\frac{1}{2}$ in KLN 
and a larger value used for $\alpha_s$, namely  $\alpha_s (Q^2) = 0.6$
at $Q^2 = 2~ GeV^2$,
allows them to get larger values of $Q_s^2$,
i.e. $Q_s^2 \ge 2~ GeV^2$ for the central collisions.

Taking in the following an optimistic point of view, we  use
(\ref{aver}) with a multiplicative $K $ factor of order $O(1)$,
explicitely  $K \simeq 1.6$,
such that for $\vec{b} = 0$,
${\bar{Q_s^2}} (\vec{b} =0) = 1 ~GeV^2$. 
Our use of $K$ is to a large extent cosmetic. This factor only appears 
in calculating $Q_s^2$, but does not change (\ref{Nard0}) or
(\ref{Nard1}). Such a factor will affect the average transverse momentum
per produced gluon but not the total number of produced gluons.
The difference between  ${\bar{Q_s}} (\vec{b} =0) = 1 ~GeV$
and ${\bar{Q_s}} (\vec{b} =0) = 0.8 ~GeV$, corresponding to
$K = 1.6 $ and $K = 1$, respectively, has little effect on
the equilibration temperature $T_{eq}$ and time $\tau_{eq}$
quoted later in this note.

viii) Finally, there is an equality assumed 
between the number of partons in the final state
and the number of observed hadrons ("parton-hadron duality"
\cite{duality}).

\vspace{0.35cm}

Let us now turn to the ``bottom-up'' scenario 
\cite{Baier1,Son,Mueller3}, which in contrast
to the KLN \cite{Nardi,Levin,Kharzeev} prescription,
does not relate the multiplicities  to the initial condition only,
but also to the way gluons are thermalized.

\begin{figure}
\centering
\epsfig{
bbllx=30,bblly=170,bburx=560,bbury=600,
file=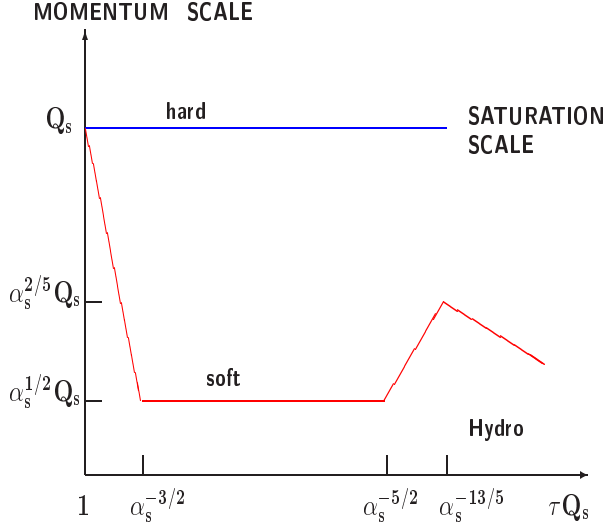,width=100mm,height=100mm}
\caption{
\label{fig:pict}
Characteristic momentum scales for the "bottom-up" scenario.}
\end{figure}

In the framework of perturbative QCD  
the time evolution of the gluonic system, when described by a non-linear
Boltzmann equation based on  $2 \leftrightarrow 2$  processes, 
a relatively long time ($\sim Q_s^{-1}~\exp(\alpha_s^{-1/2} )$) is required
for the approach to kinetic equilibration \cite{Mueller,Serreau}.
Fast and efficient thermalization occurs when inelastic
processes, namely gluon splittings 
$2 \leftrightarrow 3$ are taken into account, and
kinetic equilibration occurs much faster at times 
$\sim \alpha_s^{-13/5} Q_s^{-1}$ \cite{Baier1,Son,Mueller3}.
In this  "bottom-up" scenario the time evolution of the system 
proceeds through several regimes (Fig.~\ref{fig:pict}), with 
$Q_s\tau\sim\alpha_s^{-3/2},\alpha_s^{-5/2} $ and $\alpha_s^{-13/5}$.

The difference between the "bottom-up" and the KLN picture
is schematically illustrated in this Fig.~\ref{fig:pict}.
Under the KLN assumption the hard gluons (on the momentum scale $Q_s$)
are conserved in number, i.e. $n_{hard} \tau = const$, and they finally
 hadronize, after passing through a hydrodynamical stage.
The "bottom-up" scenario is characterized by the fact that hard gluons 
are degrading, soft ones are formed and start to dominate the system. 
As a result  the interactions of gluons in this kinetic scenario
modify strongly the initial gluon spectrum.
The gluons are  redistributed and thermalizing, such that a
quark gluon plasma  is formed.
The number of gluons, together with the entropy
\cite{Mueller3}, is increasing with proper time $\tt$,
such that the ratio is,
\begin{equation}
R = [n_{soft}(\tt) (Q_s \tt)] \vert_{      \tt_{eq}}~
/ ~ [n_{hard}(\tt) (Q_s \tt)] \vert_{\tt_0} \sim \alpha_s^{-2/5} \gg 1 \, .
\label{ratio}
\end{equation}
The following processes are expected to be present
 (at RHIC and higher 
energies in 
the central region of pseudorapidity $\eta \le 1$):  
at $Q_s \tau \ge 1$ saturated hard gluons $\longrightarrow$ elastic 
scatterings and branching/production of soft gluons $\longrightarrow$ at 
$Q_s \tau |_{eq} \sim \alpha_s^{-13/5}$ thermalization of these soft gluons 
with temperature
$T_{eq} \sim \alpha_s^{2/5} Q_s \longrightarrow$ hydrodynamic expansion 
$\longrightarrow$ hadronization (Fig.~\ref{fig:pict}).


\noindent
In more detail the parametrically estimated time scales are
determined as follows \cite{Baier1,Son,Mueller3}.

\noindent
{ Early times $1 \ll Q_s\tau \ll \alpha_s^{-3/2}$: }

At the earliest time, $\tau\sim Q_s^{-1}$, gluons, i.e. hard gluons, 
have typical large transverse momentum of order $Q_s$ and
occupation number  of order $1/\alpha_s$.
Later on
gluons with smaller momenta, but still larger than $\Lambda_{QCD}$
 will be produced (nevertheless, they are denoted as soft gluons).
The density of hard gluons $n_{hard}$ decreases with time due
to the one-dimensional expansion.
\noindent
Gluons interact by elastic scatterings
at small angle, with exchange momentum $\ll Q_s$.
The typical occupation number 
 is large until $Q_s\tau\sim\alpha_s^{-3/2}$,
when it becomes of $O(1)$.

\noindent
This regime is the transition region from the non-linear
classical gluon field to the one where the transport
description by Boltzmann equations should become applicable. 

\medskip
\noindent
{ Times $\alpha_s^{-3/2} \ll Q_s\tau \ll \alpha_s^{-5/2}$: }

Inelastic scatterings 
 produce (soft) gluons $n_{soft}$ with
characteristic  momentum estimated to be of order 
$\alpha_s^{1/2} Q_s  > \Lambda_{QCD}$ (Fig.~\ref{fig:pict}), 
namely via  $hard + hard \rightarrow hard + hard + soft$.
The number of these soft gluons becomes comparable to that
of hard ones at $Q_s\tau \sim \alpha_s^{-5/2}$,
 namely $n_{soft} \sim n_{hard}$.

\vspace{0.4cm}

\vspace{0.6cm}

\medskip
\noindent
{Times $Q_s \tt \gg \alpha_s^{-5/2}$: }

After $Q_s\tau\sim\alpha_s^{-5/2}$ most gluons are soft,
 $n_{soft}\gg n_{hard}$;
they achieve thermal equilibration amongst themselves.
Although the  whole system is still not in thermal equilibrium,
the soft part is characterized by the temperature $T$, $n_{soft} \sim T^3$.
The  few hard gluons collide with the soft ones of the thermal bath and
constantly loose energy to the latter. 
A hard gluon  emits one with a softer
energy, which
splits into  gluons with comparable momenta.
The products of this branching quickly cascade further, giving all their
energy to the thermal bath.
This  increases the temperature in the thermal bath found to be,
%
\begin{equation}
  T = c_T~ \alpha_s^3 Q_s^2 \tau \, ,
\end{equation}
i.e. it  increases linearly with time,
even when the system is expanding, due to the hard gluons which serve
as an energy source.
$c_T$ is a numerical constant to be discussed later.
One can verify a posteriori that soft gluons are indeed equilibrated due to
many interactions: the size of the expanding system of $O(\tau)$ is
indeed much larger than the mean free path of the soft gluons,
i.e.
\begin{equation}
\tau/\lambda_{soft} \sim \tau n_{soft}(\tau) \sigma 
 \sim  c_T~  \alpha_s^5 (Q_s \tau )^2 \gg 1 \, ,
\end{equation}
where the cross section is estimated as $\sigma \sim \alpha_s^2 /T^2$.

In the following, in order to provide  predictions
for particle multiplicities in this scenario 
we go as much as possible beyond the parametric estimates,
ending up with consistency arguments which specify the allowed range
of parameters.

First we start by considering the constant $c_T$ which may be written
in terms of the parameter $c$ as
\cite{Baier1},
\begin{equation}
c_T \simeq \frac{15}{8 \pi^5} c N_c^3 \,  \simeq 0.16~c \, .
\label{cT}
\end{equation}
The linear growth of $T$ as shown in Fig.~\ref{fig:pict} 
terminates, when the hard gluons loose all of
their energy, i.e.  when

\begin{equation}
\tau = \tt_{eq}
= c_{eq}~ \alpha_s^{-13/5} Q_s^{-1} \,
\label{final} ,
\end{equation}
where the parameter $c_{eq}$ is unknown for the moment,
although in principle it could be calculated in the ``bottom-up'' framework.
The temperature achieves a maximal value, i.e. the equilibration temperature, 
which is expressed by
\begin{equation}
T_{max} = T_{eq} =0.16~ c~c_{eq}~ \alpha_s^{2/5}(Q_s^2)~Q_s  \, .
\end{equation}
Subsequently the temperature decreases as $\tau^{-1/3}$ 
\cite{Bjorken}, such that $\tau n_{soft} (\tau) = const$.

We now derive the charged hadron multiplicity from
the number density of the equilibrated soft gluons,
\begin{equation}
n_{soft} (\tau_{eq}) = 2 (N_c^2 -1) \frac{\zeta(3)}{\pi^2} T_{eq}^3 \, ,
\end{equation}
which is much larger than $n_{hard}$ in (\ref{dens}) used by KLN.
One finds for the ratio $R$ defined in (\ref{ratio}),
\begin{eqnarray}
 R  = && 8 \zeta(3) N_c~ \frac{c_T^3}{c} ~\alpha_s^{10} (Q_s^2)~ (Q_s \tau)^4
\Bigg|_{\tau_{eq}}  \nonumber \\
\simeq && 0.13~ c^2~ c_{eq}^4 \alpha_s^{-2/5}(Q_s^2) \, .
\label{rat}
\end{eqnarray}

\noindent
For the charged hadron multiplicity the result
for $\sqrt{s} = 130 ~GeV$ and the most central collisions, is 
\begin{eqnarray}
\langle \frac{2}{N_{part}} \frac{d N_{ch}}{d\eta} \rangle
&& \simeq \frac{R~c}{3}~ \ln \frac{\bar{Q^2_s} }{\Lambda^2_{QCD}} 
\nonumber \\
&& \simeq 0.04~ c^3 ~ c_{eq}^4
\left[ \ln \frac{\bar{Q^2_s} }{\Lambda^2_{QCD}} \right]^{7/5} \, , 
\label{Nard1}
\end{eqnarray}
which replaces (\ref{Nard0}).

\vspace{0.5cm}

In order to compare the predicted charged hadron multiplicity,
(\ref{Nard0}) and (\ref{Nard1}),
with RHIC data we use as a reference the result by 
 the PHOBOS Collaboration \cite{Back},
namely $3.24 \pm 0.1 (stat) \pm 0.25 (syst)$, which is
in good agreement within errors  with the
experimental measurements of the other collaborations at RHIC.
This is best seen by comparing the  values for
${d N_{ch}}/{d\eta}$ at midrapidity
 at $\sqrt{s} = 130~GeV$: 
 ${d N_{ch}}/{d\eta} = 555 \pm 12 (stat) \pm 35 (syst)$ \cite{Back},
 $609 \pm 1 (stat) \pm 37 (syst)$ \cite{Adcox},
 $549 \pm 1 (stat) \pm 35 (syst)$ \cite{Bearden},
 $567 \pm 1 (stat) \pm 38 (syst)$ \cite{Adler}, respectively. 

\noindent
As reference value we take
\begin{equation}\label{norm}
\langle \frac{2}{N_{part}} \frac{d N_{ch}}{d\eta} \rangle  
= 3.24 \, .
\end{equation}

First we note that in the KLN approach agreement with experimental data
can only be achieved with the value $c \simeq 3$, which is
different from the current  numerical estimate 
quoted before, $c \simeq 0.5$.

Turning to the ``bottom-up'' approach, the experimental value (\ref{norm}) 
meets the theoretical expectation (\ref{Nard1}) 
for $R~c \simeq 3$, or equivalently
\begin{equation}\label{ceq}
c_{eq} \simeq \frac{2.0}{c^{3/4}} \, .
\end{equation}
This relation is to be confronted with the consistency requirement for
the "bottom-up" scenario , that the ratio $R$ (\ref{rat}) be larger than $2$,
implying
\begin{equation}\label{cc}
 c^2~c_{eq}^4 \ge 10, \, \, \, \,  i.e.\, \, \,  c_{eq} \ge \frac{1.8}{\sqrt{c}} \, .
\end{equation}

\noindent
It is not too difficult to see that  (\ref{ceq}, \ref{cc}) constrain
the two parameters:
\begin{equation}\label{constr}
 c~ \le~ 1.5 \, \, \,  and  \, \, \,  c_{eq} ~\ge~ 1.5  \, .
\end{equation}

\noindent
This in turn allows us to infer the actual properties of the medium,
especially to answer tentatively the question of the formation of
the equilibrated plasma.
We do this via discussing the temperature $T_{eq}$ and the equilibration time $\tau_{eq}$.
In order that the quark gluon plasma is produced, $T_{eq}$
should be bigger than the phase transition
temperature $T_{deconf}$,
which is of order $T_{deconf} =  173 \pm 8 ~MeV$
and $154 \pm 8 ~MeV$ for 2 and 3 flavour QCD, respectively \cite{Karsch},
i.e.
\begin{equation}
 T_{eq} \ge  T_{deconf}  \, .
\label{cond3}
\end{equation}
This constrains
\begin{equation}
 c~ c_{eq} >  1.3, \, \,   \,  or \, \,  c > 0.2  \, .
\label{contemp}
\end{equation}

\noindent
Finally, we may correlatively discuss $\tau_{eq}$.
Under the condition (\ref{constr}) one finds (for central collisions)
\begin{equation}
 \tau_{eq} ~ \ge 2.6 ~ fm \sim 1/2 ~R_{Au} \, ,
\label{eqtime}
\end{equation}
which is much  bigger than the current
 estimate of $\simeq 0.7 ~ fm$ \cite{McLerran}
{\footnote{This short 
equilibration time  is required in order to describe the
hadron spectra for different particles as measured at RHIC by
hydrodynamic calculations, assuming the existence of the quark gluon plasma
\cite{Heinz} .}}.
Within the uncertainties inherent to these estimates
the formation of the equilibrated plasma may indeed be realized.

One may consider the energy dependence at RHIC energies
of the charged multiplicities.
Following \cite{Kharzeev}
it is controlled by
 the  energy dependence of the saturation scale,
i.e. $Q^2_s (s)/Q^2_s (s_0) = (s/s_0)^{\lambda /2}$
with $\lambda = 0.25$ \cite{Golec}. As discussed we  choose as the reference value
$\bar{Q^2_s} (s_0) = 1~GeV^2$ at $\sqrt{s_0} = 130~GeV$.
For the most central collisions the expression for the energy dependence
of the pseudorapidity density of charged particles at midrapidity
reads in the "bottom-up" scenario,
\begin{equation}\label{energyb}
\langle \frac{2}{N_{part}} \frac{d N_{ch}}{d\eta} \rangle  
\simeq 0.64 \left( \frac{\sqrt s }{\sqrt s_0}\right)^{\lambda} \,
\left[\ln  \frac{\bar{Q^2_s} (s)}{\Lambda_{QCD}^2} \right]^{7/5}   
\, .
\end{equation}

\noindent
Based on (\ref{energyb}) the multiplicity at $\sqrt{s} = 200~GeV$ 
is obtained:
\noindent
\begin{equation}\label{edep}
\langle \frac{2}{N_{part}} \frac{dN_{ch}}{d\eta} \rangle
= 3.84  \,  , 
\end{equation}
compared to the value given by the  PHOBOS Collaboration \cite{Back},
\begin{equation}\label{PHdat}
\langle \frac{2}{N_{part}} \frac{dN_{ch}}{d\eta} \rangle 
= 3.78 \pm 0.25 ~(syst) \, . 
\end{equation} 

\begin{figure}
\centering
\epsfig{bbllx=50,bblly=200,bburx=510,bbury=615,
file=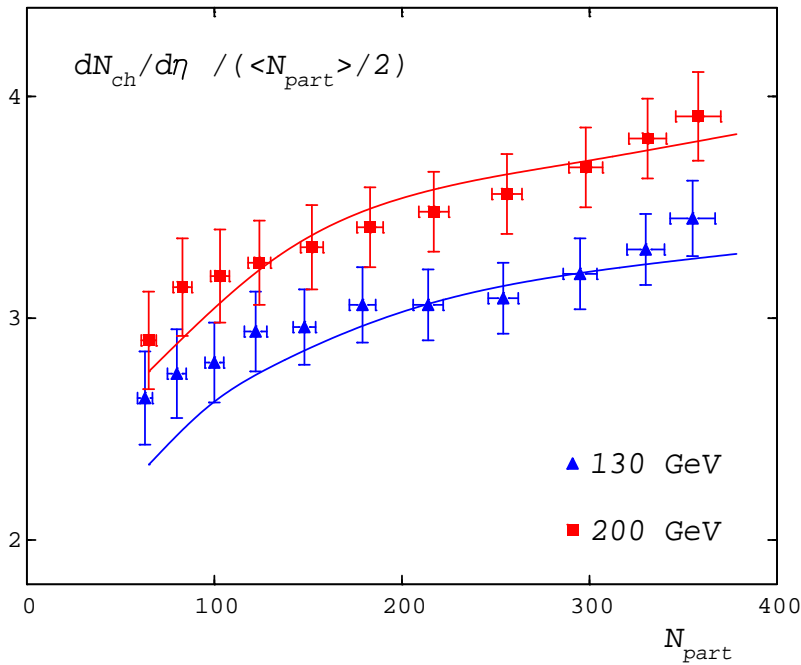, width=110mm}
\caption{
\label{fig:phob}
The scaled pseudorapidity density
for $\vert \eta \vert < 1$
as a function of $N_{part}$ at $\sqrt{s} = 130~GeV$  and $200 ~GeV$.
Data are from the PHOBOS Collaboration, and
curves from the "bottom-up" scenario, as described in the text.}
\end{figure}

\vspace{0.35cm}
Finally we consider the centrality dependence, i.e. 
 the dependence of $d N_{ch}/ d\eta$ as a function of $N_{part}$.
In Fig.~\ref{fig:phob} the  comparison of the  "bottom-up" expectation as
derived from (\ref{energyb})
 with data from the  PHOBOS Collaboration
\cite{Back} at $\sqrt{s} = 130~GeV$ and at $\sqrt{s} = 200~GeV$
 at RHIC is shown. For this comparison
 $\bar{Q^2_s} (\vec{b}) $ is calculated from (\ref{aver}) as a function of $N_{part}$,
using $\rho_{part}(\vec{b})$ given 
in Table 2 of \cite{Nardi}, together with the scaling relation
for the energy dependence of $Q_s$ quoted above \cite{Golec}. 
One has to note, that for $N_{part} < 100$ the values of $Q_s^2$ are
becoming smaller than $0.6~GeV^2$.
The  structure of the shape of $d N_{ch}/ d\eta$ as a function
of $N_{part}$ seen in the data (Fig.~\ref{fig:phob}) could be
attributed, cf. with (\ref{spect}), to  details of $x G(x, Q_s^2)$ at small $x$ as a function of $Q_s^2$.

\vspace{0.5cm}
In summary, the description provided by the ``bottom-up'' scenario
is in agreement with RHIC data,
provided the parameters $c$ and $c_{eq}$, which are not determined
in the picture, lie in a given, limited
range.
In particular for  $c \simeq 1$ and $c_{eq} \simeq 2$
the picture looks both reasonable and attractive.
For these values the results (for the most central collisions)
 for $R, T_{eq}$ and $\tau_{eq}$
are $3,~  230~MeV$ and $3.6~fm$, respectively, for $Q_s = 1~GeV$;
for $Q_s = 0.8~GeV$, i.e. for $K=1$, the values are $R=3, T_{eq} = 210 ~MeV$
and $\tau_{eq} = 3.2 ~fm$.
However,  the picture probably does not make much sense for a
small value of $c \simeq  1/2$, which is currently favoured
by numerical calculations of classical field equations,
because the ratio $R$ turns out to become $R \simeq 6 $, which 
is too much inelasticity when $Q_s = 1~GeV$. This discrepancy
deserves further examination.

On the other hand, in order to accomodate the  present RHIC data,
  the KLN description  requires $c \simeq 3$.
For the moment, due to a number of ambiguities, it is not yet clear
whether this large value is compatible with the various constraints 
of the saturation picture.

The ``bottom-up'' scenario is likely to provide a more convincing agreement with data.
However, this analysis will have to be supplemented by further ingredients, 
e.g. the calculation of $c_{eq}$, a more precise estimate of $c$,
before one can finally claim a significant agreement with data.

\vspace{0.5cm} 
\noindent 
Discussions with D.~Kharzeev, M.~Nardi, K.~Redlich and R.~Venugopalan 
are kindly acknowledged.
RB acknowledges support, in part, by DFG, project FOR 339/2-1.
The work of AHM is supported, in part, by a DOE Grant.
The work of DTS is supported, in part, by 
a DOE Grant No.\ DOE-ER-41132 and by the Alfred P.~Sloan Foundation.
This research was supported in part by the National Science Foundation
under Grant No. PHY99-07949; RB and DS thank ITP, UCSB,
for kind hospitality, and for the facilities during  
the completion of this work.

\end{document}